\documentclass[prl,superscriptaddress,twocolumn]{revtex4}
\usepackage{amssymb}
\usepackage{amsmath}\usepackage{graphicx}
\usepackage{mathrsfs}
\newcommand{\ket}[1]{|#1\rangle}

\newcommand{\Tr}{\textrm{Tr}}
\renewcommand{\vec}[1]{\boldsymbol{#1}}
\def\map#1{{\mathscr{#1}}}
\begin{document}
\title{The Quantum Cocktail Party}

\author{G. M. D'Ariano} 

\affiliation{QUIT Group,
  Dipartimento di Fisica ``A. Volta'', via Bassi 6, I-27100 Pavia,
  Italy and CNISM.} 

\author{R. Demkowicz-Dobrza\'nski}

\affiliation{Center for Theoretical Physics of the Polish Academy of
  Sciences, Aleja Lotnik\'ow 32/44, 02-668 Warszawa, Poland.} 

\author{P. Perinotti} 

\affiliation{QUIT Group,
  Dipartimento di Fisica ``A. Volta'', via Bassi 6, I-27100 Pavia,
  Italy and CNISM.} 

\author{M. F. Sacchi}

\affiliation{QUIT Group,
  Dipartimento di Fisica ``A. Volta'', via Bassi 6, I-27100 Pavia,
  Italy and CNISM.}

\affiliation{CNR - Istituto Nazionale per la Fisica della Materia, 
Unit\`a di Pavia, Italy.}  

\begin{abstract}
  We consider the problem of decorrelating states of coupled quantum
  systems. The decorrelation can be seen as separation of quantum
  signals, in analogy to the classical problem of signal-separation
  rising in the so-called cocktail-party context. The separation of
  signals cannot be achieved perfectly, and we analyse the optimal
  decorrelation map in terms of added noise in the local separated
  states. Analytical results can be obtained both in the case of
  two-level quantum systems and for Gaussian states of harmonic
  oscillators.
\end{abstract}

\maketitle 
In its digital form, {\em information} is perfectly copy-able and
broadcastable at will.  In its analog form of everyday life, however,
information often comes mixed-up. This is the case, for example, when
we join a cocktail party, and we hear two people speaking
simultaneously: their voices come together in one signal to our ears.
Our brain is easily able to ''tune'' to one voice and ignore the
other, and, sometimes to even grasp both of them. If, however, we want
to digitalise the two speeches separately, we need to de-mix the two
voices, and this is generally a hard task for a neural-network
software, a problem which is indeed commonly known as {\em the
  cocktail party problem} \cite{cock}. In Quantum Mechanics we have a
similar situation for the {\em quantum information}. We known that
quantum information cannot be copied or broadcast exactly, due to the
no-cloning theorem \cite{Wootters82} (which asserts the impossibility
of making exact copies of an unknown quantum state drawn from a non
orthogonal set). Such a limitation is actually very valuable for
quantum cryptography, as it forbids an eavesdropper from creating
copies of a transmitted quantum cryptographic key. In the presence of
noise, however, (i.  e. when transmitting "mixed" states), it can
happen that we are able to increase the number of copies of the same
state if we start with sufficiently many identical originals. Indeed,
it is even possible to {\em purify} in such broadcasting process---the
so-called {\em super-broadcasting} \cite{our}.  Clearly, the increased
number of copies cannot augment the available information about the
original input state, and this is actually due to the fact that the
final copies are not statistically independent, and the correlations
between them influence the extractable information \cite{estcor}.  It
is now natural to ask if we can remove such correlations and make them
independent again, a process which is a quantum analog of the cocktail
party problem. Clearly, such quantum un-mixing or de-correlating
cannot be done exactly, otherwise we would increase the information on
the state. However, we will show here that we can achieve perfect
de-correlation at expense of some more noise in each copy.\par

In the typical cocktail party scenario we have two microphones in the
same room at different locations. If we denote the amplitude of a
sound wave emitted by two people by $\alpha(t)$, $\beta(t)$
respectively, then microphones will in general record a linear
combination of this messages (for simplification, we neglect the
possible delays in the time arrival to different receivers from
different sources):
\begin{align}
x(t)&=C_{11} \alpha(t) + C_{12} \beta(t) \\
y(t)&=C_{21} \alpha(t) + C_{22} \beta(t)
\end{align}
where $C_{ij}$ are parameters which depend on the microphone
sensitivities and on their distance from the speakers, and $x(t)$,
$y(t)$ are the recorded signals. Amazingly, even if the parameters
$C_{ij}$ are unknown and signals $\alpha(t)$, $\beta(t)$ do not have
any distinctive feature (e.g.  different frequency band), the
separation of original signals is still possible, under the sole
assumption that original signals where uncorrelated (and the
additional technical assumption that the probability distributions of
the signals amplitude at different times were not Gaussian).  A way to
achieve the un-mixing task is by the so called \emph{independent
  component analysis} (ICA), which uses the fact that the probability
distribution of a sum of independent random variables is ''more
Gaussian'' than the probability distribution of the variables
themselves. This strategy is sometimes called the \emph{blind
  independent component analysis}, as we know neither the signal
probability distribution nor the mixing parameters $C_{ij}$. After a
successful application of the above strategy one is left with
independent signals $\alpha^\prime(t)$ and $\beta^\prime(t)$, which in
the ideal case differ from the input signals $\alpha(t)$ and
$\beta(t)$ by a scaling factor (in reality there is always some
noise), and are uncorrelated. Mathematically the de-correlation task
can be stated in term of factorization of the conditional probability
distributions as follows
\begin{equation}\label{decorprob}
p_{AB}(\alpha^\prime, \beta^\prime | \alpha, \beta) =
p_A(\alpha^\prime| \alpha)p_B(\beta^\prime | \beta).
\end{equation}
\begin{figure}[ht]
  \includegraphics[width= 0.5\textwidth]{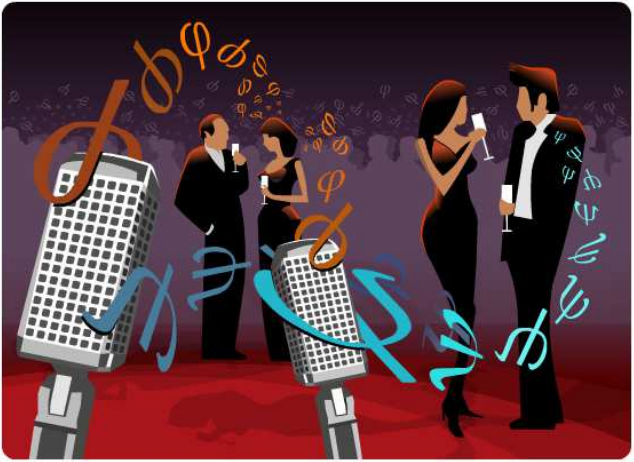}
  \caption{When we join a cocktail party and we hear two people
    speaking simultaneously, their voices come together in a single
    signal to our ears (which we can simulate by two microphones).  If
    we want to digitalise the two speeches separately, we then need to
    de-mix the two voices, and this is generally a hard task for a
    neural-network software, a problem which is indeed commonly known
    as {\em the cocktail party problem}. In Quantum Mechanics we can
    also consider a situation of de-correlating two signals, but in
    this case the signals are not classical, and are encoded using
    quantum states. [Picture courtesy by Tomasz
    Szkodzi\'nski].}\label{f:cocktail}
\end{figure}
\par A quantum strict analog of the problem can be formulated as
follows.  Assume we have a bipartite quantum system (e.g. two qubits,
two quantum modes of electromagnetic field, etc.) initially in a state
$\ket{0}\otimes \ket{0}$ (or more generally in some mixed state
$\rho_{AB}$). The signal is encoded using unitary operations $U_A(t)$,
$U_B(t)$ acting locally at time $t$ on subsystems $A$ and $B$,
respectively. The communication of quantum signals will amount to
sending the states $[U_A(t)\otimes U_B(t)]\ket{0}\otimes \ket{0}$ at
different times $t$, each time rotated by a different pair of unitary
matrices $U_A(t)$ and $U_B(t)$, depending on the quantum message
intended to be transmitted.  After this encoding, the system passes
through the environment which causes the two signals to be mixed in
analogy to classical mixing of signals in microphones. This mixing can
be represented by a unitary operation $V$ that entangles both qubits
with the environment state $\ket{E}$ as follows
\begin{equation}
\ket{\psi(t)}_{ABE}=V(U_A(t)\otimes U_B(t)\otimes I) \ket{0} \otimes \ket{0}\otimes \ket{E}.
\end{equation}
The analog of the classical cocktail-party problem would be now to
determine the ``signals'' $U_A(t)$ and $U_B(t)$---or the state
$[U_A(t)\otimes U_B(t)]\ket{0}\otimes \ket{0}$---from the output state
of $AB$ only, without even knowing the interaction with the
environment $V$: this would be a strict quantum analog of \emph{blind
  independent component separation}. In this sense we would
de-correlate the signals $U_A(t)$ and $U_B(t)$.  This quantum version
of the cocktail-party problem is much harder than its classical
counterpart, for many reasons, including the no-cloning theorem, which
forbids to determine the output state from a single copy: an
approximate solution, if possible, would need at least some additional
assumptions about the time self-correlation of each separate signal,
along with the aid of a quantum memory to store the whole
time-sequence of output states of $AB$ and a full joint measurement on
the whole sequence.

We pose here a simpler, but a closely related problem of
de-correlating two quantum signals, in the scenario where the signals
$U_A$, $U_B$ are encoded on a correlated state $\rho_{AB}$ as: $U_A
\otimes U_B \rho_{AB} U^\dagger_A \otimes U^\dagger _B$, but no
additional mixing operation $V$ is applied.  We want to de-correlate
the received state, and the desired result is two completely
uncorrelated systems $A$ and $B$, each one in a state that carries
information about the signals $U_A$ and $U_B$, respectively.

Therefore, according to the above scenario, let $\rho_{AB}$ be a
density matrix of two (generally correlated) quantum systems. The
hardest case will be when the two systems $A$ and $B$ are identical,
and the state $\rho_{AB}$ doesn't change under permutation of them.
The information is encoded on the state $\rho_{AB}$ via the local
unitary transformations as follows
\begin{equation}
\rho_{AB}(\alpha,\beta)\doteq U_A(\alpha)\otimes U_B(\beta) \rho_{AB} U_A^\dagger(\alpha)\otimes U_B^\dagger(\beta),
\end{equation}
$\alpha$ and $\beta$ denoting random variables.  The de-correlating quantum transformation $\map{D}$
we are seeking should act as follows:
\begin{equation}
\rho_{AB}(\alpha,\beta)\longrightarrow\tilde{\rho}_A(\alpha) \otimes \tilde{\rho}_B(\beta)
\end{equation}
with $\tilde\rho_A(\alpha)\doteq U_A(\alpha)\tilde\rho_A
U_A^\dagger(\alpha)$, and $\tilde\rho_B(\beta)\doteq
U_B(\beta)\tilde\rho_B U_B^\dagger(\beta)$. This means that the map
acts {\em covariantly} with respect to the action of
$U_A(\alpha)\otimes U_B(\beta)$. The output state is uncorrelated, and
we want the matrices $\tilde{\rho}_A(\alpha)$ and
$\tilde{\rho}_B(\beta)$ to contain as little noise as possible, namely
they will carry the same signal, but possibly with higher noise.  In
other words, we want the states $\tilde\rho_A(\alpha)$ and
$\tilde\rho_B(\beta)$ to be as close as possible to the input marginal
states $\rho_A(\alpha)=\Tr_B[\rho_{AB}(\alpha,\beta)]$,
$\rho_B(\beta)=\Tr_A[\rho_{AB}(\alpha,\beta)]$, respectively.
\par At the output the two classical signals $\alpha$ and $\beta$
encoded on the joint state $\rho_{AB}(\alpha,\beta)$ are recovered by
separate identical measurements on systems $A$ and $B$, yielding the
probability distribution
\begin{equation}
  p_{AB}(\alpha^\prime,\beta^\prime|\alpha,\beta)=\Tr[\Pi(\alpha^\prime)\otimes
  \Pi(\beta^\prime) \rho_{AB}(\alpha,\beta)],
\end{equation}
where $\Pi(\alpha)$ and $\Pi(\beta)$ are positive operators describing
the local measurements on $A$ and $B$, fulfilling the normalization
condition $\int \textrm{d}\alpha^\prime \Pi(\alpha^\prime) =\int
\textrm{d}\beta^\prime \Pi(\beta^\prime) = \openone$.

If instead we first apply the de-correlation operation $\map{D}$, and
then perform the measurements we get the probability distribution
\begin{equation}
\begin{split}
  p^\map{D}_{AB}(\alpha^\prime, \beta^\prime | \alpha,
  \beta)=&\Tr[\Pi(\alpha^\prime)\otimes \Pi(\beta^\prime)
  \tilde{\rho}_{A}(\alpha) \otimes \tilde{\rho}_B(\beta)]\\=&
  p_A(\alpha^\prime|\alpha)p_B(\beta^\prime|\beta),
\end{split}
\end{equation}
achieving the solution of the cocktail party problem as in Eq.
(\ref{decorprob}). We want to stress that the de-correlated
probabilities $p_A(\alpha^\prime|\alpha)$ and
$p_B(\beta^\prime|\beta)$ will be generally more noisy than the
respective marginals of the original joint probability (indeed a
perfect de-correlation is not possible, since it would violate
linearity of quantum mechanical evolutions: see also Refs.
\cite{Terno,Mor}).  
\par Now we will show how de-correlation can be achieved in two
specific examples: on {\em qubits}, and on {\em qumodes} (the
so-called {\em continuous variables}, i.~e. quantum harmonic
oscillators).

\par
\begin{figure}
\includegraphics[width= 0.5\textwidth]{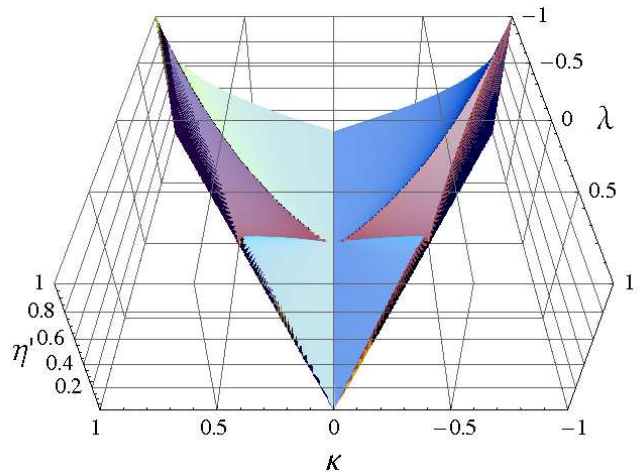}
\caption{Length $\eta'$ of the Bloch vectors of the de-correlated
  states of two qubits starting from the joint state in Eq.
  (\ref{decorrelable}). The 3D plot depicts the maximal achievable
  $\eta^\prime $ versus the parameters $\kappa$ and $\lambda$ of the
  input state.}\label{f:eta}
\end{figure}

 Consider a couple of qubits. For qubits the state is conveniently described in the Bloch
form. The information $(\alpha,\beta)$ is encoded by $U_A(\alpha)$ and $U_A(\beta)$ on the direction
of the Bloch vectors $\vec{n}_A(\alpha)$ and $\vec{n}_B(\beta)$ of the marginal states
\begin{equation}
\begin{split}
\rho_A(\alpha)=&\Tr_B[\rho_{AB}(\alpha,\beta)]=\tfrac{1}{2}(\openone + \eta\vec{n}_A(\alpha)\cdot\vec{\sigma}),
\\
\rho_B(\beta)=&\Tr_A[\rho_{AB}(\alpha,\beta)]=\tfrac{1}{2}(\openone + \eta \vec{n}_B(\beta)\cdot\vec{\sigma}),
\end{split}
\end{equation}
where $\vec\sigma=(\sigma_x,\sigma_y,\sigma_z)$ is the vector of Pauli matrices $\sigma_\alpha$. 
Covariance of the de-correlation map means that the direction of the Bloch vectors
$\vec{n}_A(\alpha)$ and $\vec{n}_B(\beta)$  should be preserved in the output states, i.~e.
\begin{equation}
\begin{split}
  & \tilde{\rho}_A(\alpha)=\frac{1}{2}(\openone + \eta^\prime
  \vec{n}_A(\alpha)\cdot\vec{\sigma}),\\
  & \tilde{\rho}_B(\beta)=\frac{1}{2}(\openone +\eta^\prime
  \vec{n}_B(\beta)\cdot\vec{\sigma}),
\end{split}
\end{equation}
namely only the length of the Bloch vector (i.~e. the purity of the
state) is changed $\eta\to\eta'$. The fact the output states are more
noisy corresponds to a reduced length of the Bloch vector
$\eta'<\eta$. The directions of the Bloch vectors $\vec n_A(\alpha)$
and $\vec n_A(\beta)$ are completely arbitrary. The optimal
de-correlation map will maximize the length $\eta'$ of the Bloch
vector, namely it will produce the highest purity of de-correlated
states. It can be shown 
that the general form of a
two-qubit channel $\map{D}$ covariant under $U_A(\alpha) \otimes
U_B(\beta)$ and invariant under permutations of the two qubits can be
parameterized with three positive parameters only (effectively two due
to normalization)
\begin{equation}
\map{D}(\rho_{AB})= a \rho_{AB} + b \map{D}_1(\rho_{AB})+c
\map{D}_2(\rho_AB),
\end{equation}
where $\map{D}_1$ and $\map{D}_2$ are given by
\begin{align}
\map{D}_1(\rho_{AB})=&\tfrac{1}{3}\left(\openone \otimes \rho_B +
\rho_A \otimes \openone - \rho_{AB} \right)\\
\map{D}_2(\rho_{AB})=&\tfrac{1}{9}\left(4 \openone \otimes \openone +
\rho_{AB}- 2 \openone \otimes \rho_B -2 \rho_A \otimes
\openone\right)
\end{align}
and the trace preserving condition gives $a+b+c=1$. This is a very
restricted set of operations, which is due to the fact that the
covariance condition is very strong. As a consequence a generic joint
state $\rho_{AB}$ cannot be de-correlated (of course, apart
from the trivial de-correlation to a maximally mixed state), and the
states for which de-correlation is possible have the form 
\begin{equation}\label{decorrelable}
\rho_{AB}=\tfrac{1}{4}\left[\openone \otimes \openone + \kappa
(\sigma_z \otimes \openone + \openone \otimes \sigma_z) + \lambda
\sigma_z \otimes \sigma_z \right].
\end{equation}
We emphasize that for a generic state $\rho_{AB}$ one can reduce
correlations, but only for states of the form (\ref{decorrelable}) the
correlations can be completely removed. The noise of the de-correlated
states depends on parameters $\kappa$ and $\lambda$ as depicted in
Figure \ref{f:eta}.

\par We consider now the case of de-correlation for qumodes. For a
couple of qumodes in a joint state $\rho_{AB}$ the information
$(\alpha,\beta)$ (with $\alpha$ and $\beta$ complex) is encoded as
follows
\begin{equation}\label{qstate}
  D(\alpha )\otimes D(\beta)\rho_{AB} D(\alpha )^\dagger \otimes D(\beta)^\dagger,
\end{equation}
$D(z)=\exp(za^\dag-z^*a)$ for $z\in{\mathbb C}$ denoting a single-mode
displacement operator, $a$ and $a^\dag$ being the annihilation and
creation operators of the mode.  In particular, it can be shown
that it is always possible to de-correlate any joint
state of the form (\ref{qstate}), with $\rho _{AB}$ representing a
two-mode Gaussian state, namely
\begin{eqnarray}
\rho_{AB} = \frac {1}{\pi ^4}\int d^4\vec{q} \,e^{-\frac 12 \vec{q}^T\vec{M}\vec{q}}D(\vec{q})\;,
\end{eqnarray}
where $\vec{M}$ is the $4 \times 4$ (real, symmetric, and positive)
correlation matrix of the state, $\vec{q}=(q_1,q_2,q_3,q_4)$, and
$D(\vec{q})=D(q_1 +i q_2)\otimes D(q_3 +i q_4)$. The de-correlation
channel covariant under $D(\alpha)\otimes D(\beta)$ is given by
\begin{equation}\label{qdecorr}
  \map{D}(\rho)= \frac {\sqrt{\hbox{det}\vec{G}}}{(2\pi)^2}\int d^4
  \vec{x}\,e^{-\frac 12\vec{x}^T\vec{G}\vec{x}} D(\vec{x}) \rho D^\dag (\vec{x}),
\end{equation}
with suitable positive matrix $G$, and the resulting state $\map
{D}(\rho _{AB})$ is still Gaussian, with a new block-diagonal
covariance matrix $\widetilde{\vec M}$, thus corresponding to a
de-correlated state. 

A special example of Gaussian state of two qumodes is the so-called
{\em twin beam}, which can be generated in a quantum optical lab by
parametric down-conversion of vacuum. In this case $\vec M$ is given
by
\begin{eqnarray}
\vec M= \frac {1+ \lambda ^2}{1- \lambda ^2} \openone - \frac {2 \lambda }{1-
  \lambda ^2} \left ( 
\begin{array}{cc}
0 & \sigma _z \\ \sigma _z &0 
\end{array}
\right )\;,
\end{eqnarray}
with $0\leq \lambda <1$. 
The map (\ref{qdecorr}) with 
\begin{eqnarray}
\vec G= \frac {2 \lambda }{1- \lambda ^2} 
\left [\openone + \left ( 
\begin{array}{cc}
\varepsilon  & \sigma _z \\ \sigma _z & \varepsilon 
\end{array}
\right )\right ]\;,
\end{eqnarray}
and arbitrary $\varepsilon >0$, provides two de-correlated states with
$\widetilde{\vec M} = (\frac{1+ \lambda }{1- \lambda }+\varepsilon )
\openone $, which correspond to two thermal states with mean photon
number $\bar n= \frac{\lambda }{1-\lambda }+\frac {\varepsilon}{2}$
each.

\par The striking difference between the qubit and the qumode
cases is that for qubits only few states can be de-correlated, whereas
for qumodes any joint Gaussian state can be de-correlated. This is due
to the fact that the covariance group for qubits comprises all local
unitary transformations, whereas for qumodes includes only local
displacements, which is a very small subset of all possible local
unitary transformations in infinite dimension. Indeed, for the same
reason de-correlation becomes much easier when considering covariance
with respect to unitary transformations of the form $U\otimes U$
(i.~e. with the same information encoded on the quantum systems, e.~g.
the qubit Bloch vectors have the same direction, or the qumodes are
displaced in the same direction), which is actually the case when
considering broadcasted states. Covariant de-correlation of this kind
for multiple copies gives insight into the problem of how much
individual information can be preserved, while all correlations
between copies are removed. 

\acknowledgments G. M. D. is grateful to M. Raginsky for having
attracted his attention to the problem of blind source separation.
This work has been supported by Ministero Italiano dell'Universit\`a e
della Ricerca (MIUR) through FIRB (bando 2001) and PRIN 2005 and the
Polish Ministry of Scientific Research and Information Technology
under the (solicited) grant No.  PBZ-Min-008/P03/03.

\end{document}